\documentclass[aps,prb,twocolumn,showpacs,byrevtex]{revtex4}

\usepackage{times,mathptm}
\usepackage[dvips]{graphicx,color}
\newcommand{\bqa}{\begin{eqnarray*}}
\newcommand{\eqa}{\end{eqnarray*}}

\begin{document}

\title{Hydrogen adsorption-induced nanomagnetism at the Si(111)-(7$\times$7) surface}
\author{Xiao-Yan Ren$^1$, Chun-Yao Niu$^1$, Seho Yi$^2$, Shunfang Li$^1$, and Jun-Hyung Cho$^{2,*}$}
\affiliation{
$^1$ International Laboratory for Quantum Functional Materials of Henan, and School of Physics and Engineering, Zhengzhou University, Zhengzhou 450001, China\\
$^2$ Department of Physics, Research Institute for Natural Science, and HYU-HPSTAR-CIS High Pressure Research Center, Hanyang University, 222 Wangsimni-ro, Seongdong-Ku, Seoul 04763, Korea}

\date{\today}

\begin{abstract}
The creation of magnetism on non-magnetic semiconductor surfaces is of importance for the realization of spintronics devices. Especially, the coupling of electron spins within quantum nanostructures can be utilized for nanomagnetism applications. Here, we demonstrate, based on first-principles density-functional theory calculations, that the adsorption of H atoms on the Si(111)-(7$\times$7) surface induces the spin polarization of surrounding Si dangling bonds (DBs) and their spin orderings. It is revealed that the H adsorption on a rest-atom site exhibits a Jahn-Teller-like distortion that accompanies a charge transfer from the rest atom to the nearest neighboring adatoms. This charge transfer increases the local density of states of such three adatoms at the Fermi level, thereby inducing a Stoner-type instability to produce a ferrimagnetic order of adatom DBs around the adsorbed H atom. Meanwhile, the H adsorption on an adatom site cannot induce spin polarization, but, as adsorbed H atoms increase, the ferrimagnetic order of rest-atom DBs emerges through the charge transfer from rest atoms to adatoms. Our findings provide a microscopic mechanism of the H-induced spin orderings of Si DBs at the atomic scale, which paves a novel way to the design of nanoscale magnetism in the representative semiconductor surface.
\end{abstract}


\maketitle
\section{INTRODUCTION}
Nanoelectronic devices have been developed by the identification of various types of nanostructures such as quantum dots, quantum wires, and quantum wells~\cite{folsch,bohou}.  These artificially fabricated quantum structures often exhibit a number of exotic phenomena different from their bulk counterparts due to confined electrons in the low dimensions~\cite{haider,kepen}. Especially, the coupling of electron spins within such nanostructures can be utilized for the design of nanoscale spintronics devices where the functionality of individual atomic spins is geared towards storage capacity, computing speed, and energy saving~\cite{oleg,yengui,kane}. In the present study, we demonstrate that the extensively investigated Si(111)-(7$\times$7) surface exhibits the spin orderings of Si DBs (which can be treated as quantum dots)~\cite{Schofield} at the atomic scale around adsorbed H atoms, thereby providing a playground for the atomic engineering of future spintronics and quantum information devices.

Similarly, a recent scanning tunneling microscopy (STM) experiment~\cite{hector} observed that the adsorption of a single H atom on graphene induces a magnetic moment by removing one $p_z$ orbital from the delocalized ${\pi}$-bonding network, which has already been predicted by purely theoretical works~\cite{yazy}. Because of the bipartite sublattices A and B of graphene, the magnetic moment is essentially localized on the carbon sublattice opposite to the one where the H atom is chemisorbed. Such an H-induced magnetic moment exhibits a large spatial extension over several next-nearest neighbors away from the H atom~\cite{hector}. Meanwhile, the presently predicted H-induced magnetic moment on Si(111)-(7$\times$7) extends only up to the nearest neighboring adatoms or rest atoms, as discussed below. Thus, the present H-induced magnetism on Si(111)-(7$\times$7) shows a strongly localized character which would be useful for nanomagnetism applications.

Due to its complex structural and electronic properties, the Si(111)-(7$\times$7) surface has been the most extensively studied semiconductor surface over several decades. According to the Takayanagi's model~\cite{takaya}, this surface has 19 DBs per 7$\times$7 unit cell, where 12 DBs belong to the adatoms, 6 DBs to the rest atoms, and 1 DB to the Si atom at the bottom of corner hole [see Fig. 1(a)]. It has been well established that there is an electronic charge transfer from the adatoms to rest atoms~\cite{north}. Consequently, the DB states arising from the rest atoms and corner hole atoms are completely filled, while those from the adatoms are partially occupied to leave a band that crosses the Fermi level $E_{\rm F}$ [see Fig. 1(b)]. Photoemission and electron energy-loss spectroscopies measured an intriguing density of states (DOS) around $E_{\rm F}$, while STM observed different local DOS (LDOS) depending on the surface atoms [see Fig. 1(c)]~\cite{prb2005}. By employing adsorbates on Si(111)-(7$\times$7), this LDOS can be tuned to influence its structural and electronic properties. For instance, when H atoms adsorb on the rest-atom sites of the Si(111)-(7$\times$7) surface, an earlier STM experiment~\cite{prb2005} revealed that there is a reverse charge transfer from the rest atoms back to the surrounding adatoms. Although such an H-adsorbed Si(111)-(7$\times$7) surface system has so far been much studied experimentally and theoretically, most of them were related with the adsorption sites and diffusion path or the structural phase transition induced by H adsorption~\cite{vittadini,lo,fuku,morten,yurt,durr,arai}. Interestingly, using density-functional theory (DFT) calculations, Okada $et$ $al$~\cite{okada}. predicted the magnetic ordering of several DB networks on an otherwise H-covered Si(111)-(1$\times$1) surface, whereas Erwin and Himpsel~\cite{erwin} also predicted the localized spin arrays of DB electrons along the Si step edges in the Au-induced vicinal Si(111) surfaces. However, the investigation of magnetism on a realistic H-adsorbed Si(111) surface with the 7$\times$7 structure is still lacking.

\begin{figure}[htb]
\centering{ \includegraphics[width=8.0cm]{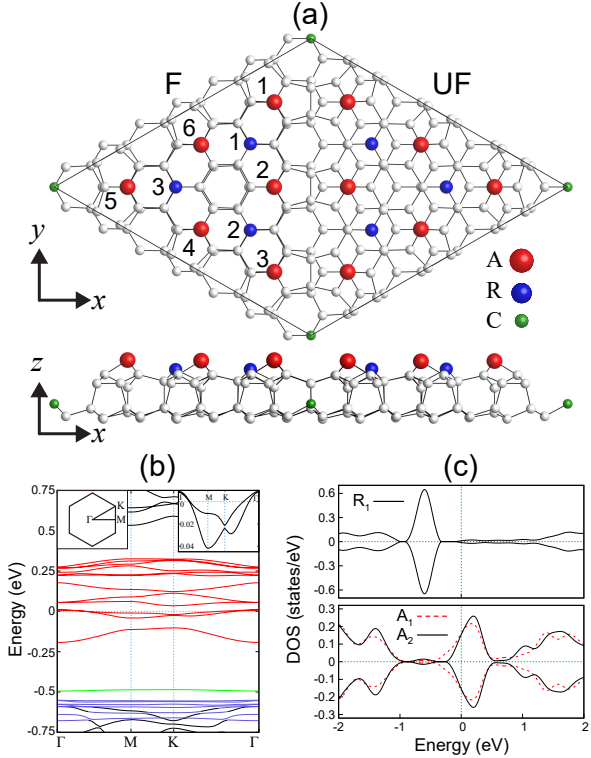} }
\caption{(Color online) (a) Top and side views of the optimized structure of the Si(111)-(7$\times$7) surface. The colored large, medium, and small circles represent the adatoms (A), rest atoms (R), and corner-hole atoms (C), respectively. F and UF represent the faulted and the unfaulted half cells, respectively. The adatoms and rest atoms in the F half cell are numbered. The calculated band structure of the Si(111)-(7$\times$7) surface is given in (b). Here, the bands originating from A, R, and C atoms are distinguished with the same colors as used in (a) and the energy zero represents the Fermi level. The surface Brillouin zone of the 7$\times$7 unit cell is drawn in the inset of (b). The LDOS obtained at R$_1$, A$_1$, and A$_2$ is given in (c). }
\end{figure}

In this paper, we take the Si(111)-(7$\times$7) surface to investigate the effect of H adsorption on possible spin orderings of Si DBs using first-principles DFT calculations. We reveal that under H adsorption the charge redistribution of DB electrons takes place differently depending on the adsorption sites. We find that the H adsorption on a rest-atom site causes a charge transfer back to its nearest neighboring adatoms, thereby leading to an increased LDOS of such three adatoms at $E_{\rm F}$. The resulting high LDOS induces the Stoner-type instability to yield the ferrimagnetic order of adatom DBs around the adsorbed H atom. On the other hand, when an H atom adsorbs on an adatom site, a charge abstraction takes place from its surrounding adatoms to the adatom, forming an H$-$Si bond without spin polarization. However, as the number of adsorbed H atoms within the 7$\times$7 unit cell increases, some rest atoms begin to participate in charge donation to the H$-$Si bond formations, therefore inducing the ferrimagnetic order of the rest-atom DBs. The present results not only elucidate the underlying mechanism of the H-induced spin orderings of DBs on the Si(111)-(7$\times$7) surface, but also open a new research area of tailoring nanomagnetism on the representative semiconductor surface by using adsorbates.

\section{COMPUTATIONAL METHODS}

The present DFT calculations were performed using the Vienna {\it ab initio} simulation package with the projector-augmented wave (PAW) method~\cite{vasp1,vasp2,paw}. For the exchange-correlation energy, we employed the generalized-gradient approximation functional of Perdew-Burke-Ernzerhof (PBE)~\cite{pbe}. The Si(111)-(7$\times$7) surface (with the optimized lattice constant 5.431 {\AA}) was modeled by a periodic slab geometry consisting of the seven-layer slab and ${\sim}$15 {\AA} of vacuum in between the slabs. Here, each Si atom in the bottom layer was passivated by one H atom. A plane-wave basis was employed with a kinetic energy cutoff of 300 eV, and the ${\bf k}$-space integration was done with 2${\times}$2 uniform meshes in the surface Brillouin zone. All atoms except the bottom two Si layers were allowed to relax along the calculated forces until all the residual force components were less than 0.01 eV/{\AA}.

\section{Results and discussion}

We begin by optimizing the atomic structure of a clean Si(111)-(7$\times$7) surface to examine its electronic band structure. Figure 1(a) shows the optimized structure of the dimer-adatom-stacking fault (DAS) model of Si(111)-(7$\times$7)~\cite{takaya}. The DAS model features the presence of 12 adatoms, 6 rest atoms, and 1 corner-hole atom, each of which offers one DB electron. The calculated band structure of Si(111)-(7$\times$7) is given in Fig. 1(b). We find that the surface states originating from 19 DB electrons of the adatoms, rest atoms and corner-hole atom are located at $-$0.18${\sim}$0.32 eV, $-$0.68${\sim}$$-$0.55 eV, and $-$0.48 eV around $E_{\rm F}$, respectively, in good agreement with previous DFT calculations and STM measurements~\cite{geun,hamer}. Therefore, the 7 surface bands arising from the rest atoms and core-hole atom are completely filled by 14 DB electrons, while the remaining 5 DB electrons occupy two and half surface bands arising from the adatoms. This occupation of surface bands indicates a charge transfer from the adatoms to the rest and corner-hole atoms. In Fig. 1(a), there are two symmetrically different types of adatoms, i.e., coner adatoms (A$_1$, A$_3$, A$_5$) and center adatoms (A$_2$, A$_4$, A$_6$) in the faulted (F) half cell. The LDOS obtained at the A$_1$ and A$_2$ adatoms exhibit similar patterns [see Fig. 1(c)], each of which is also nearly identical to that at the counterpart adatom in the unfaulted (UF) half cell (see Fig. S1 of the Supplemental Material). Therefore, each adatom is likely to have ${\sim}$5/12 DB electrons. It is noticeable that the peaks of the LDOS of the adatoms A$_1$ and A$_2$ are located just above $E_{\rm F}$, whereas those of the rest atoms (R$_1$, R$_2$, R$_3$) are identically located below $E_{\rm F}$ [see Fig. 1(c)]. By employing H adsorption on the Si(111)-(7$\times$7) surface, the LDOS peak positions of adatoms or rest atoms can be shifted toward $E_{\rm F}$, as discussed below. The resulting increased LDOS at $E_{\rm F}$ may in turn give rise to the local spin polarization derived from the Stoner instability~\cite{stoner}. In this sense, tuning the LDOS at $E_{\rm F}$ is very challenging for the realization of the spin orderings of DBs around adsorbed H atoms on the Si(111)-(7$\times$7) surface.

To explore how the LDOS changes by H adsorption on the Si(111)-(7$\times$7) surface, we first consider the adsorption of a single H atom on the rest atom R$_1$ within the F half cell, which was well known as the most stable adsorption site~\cite{lo,prb2005,vittadini}. The optimized structure using the spin-unpolarized DFT calculation shows that the height of R$_1$ (A$_1$, A$_2$, and A$_6$) is lowered (raised), relative to that (those) of the clean Si(111)-(7$\times$7) surface, by 0.41 {\AA} (0.10, 0.09, and 0.09 {\AA}). Such a Jahn-Teller-like distortion~\cite{hito} is accompanied by a charge transfer from R$_1$ to the three neighboring adatoms. This charge transfer can be confirmed by examining the LDOS of each atom. Unlike the case of the clean Si(111)-(7$\times$7) surface [see Fig. 1(c)], the LDOS peak of R$_1$ disappears upon the H adsorption [see Fig. 2(a)], but its LDOS sum below $E_{\rm F}$ keeps invariant. Further, the LDOS peaks of A$_1$ and A$_2$ (A$_6$) become closer to $E_{\rm F}$ [see Fig. 2(a)], whereas those of other adatoms remain nearly unchanged compared to the clean Si(111)-(7$\times$7) surface [see Fig. S2 of the Supplemental Material]. These results indicate that the H adsorption on the R$_1$ atom induces a local charge redistribution mostly up to the nearest neighboring adatoms. Noting that the sum of the LDOS of A$_1$ or A$_2$ (A$_6$) increases because of their shifts to higher binding energy, we can say that one excess electron obtained by H adsorption is transferred from the R$_1$ atom to its neighboring adatoms A$_1$, A$_2$, and A$_6$. Figure 2(b) shows the schematic diagram for the interaction of an adsorbed H atom with the rest atom. Due to the fully occupied DB electrons at the rest atom, formation of the Si$-$H bond leads one excess electron to occupy the antibonding state. However, since such an antibonding state is higher in energy than the surface state located at $E_{\rm F}$, it is natural that a local charge transfer from the R$_1$ atom to its surrounding adatoms A$_1$, A$_2$, and A$_6$ takes place through the above-mentioned Jahn-Teller-like distortion, lowering the electronic energy of the H/Si(111) system.

\begin{figure}[htb]
\includegraphics[width=8.0cm]{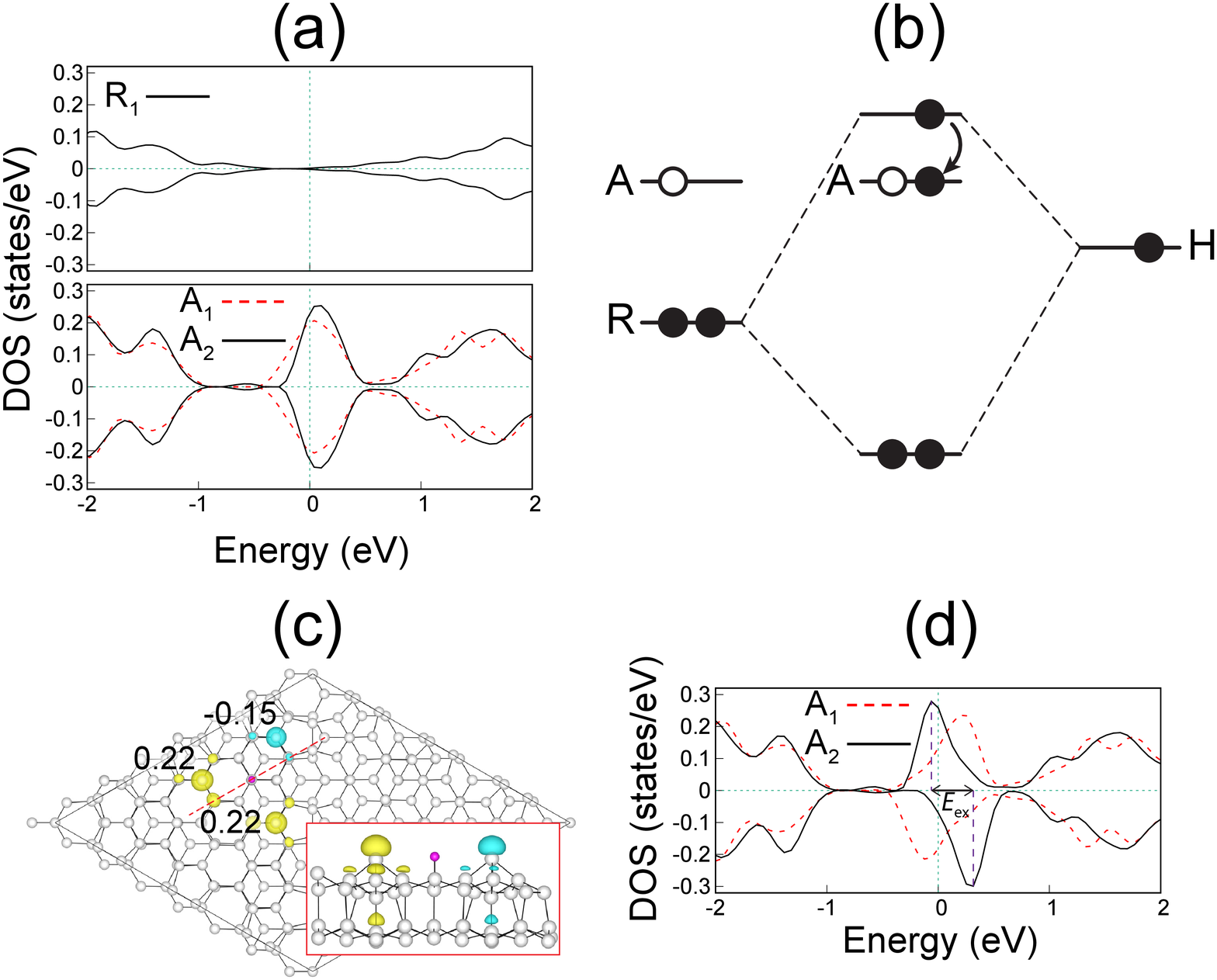}
\caption{ (Color online) (a) calculated LDOS of the R$_1$, A$_1$, and A$_2$ atoms for the adsorption of a single H atom on the rest atom R$_1$, obtained using the spin-unpolarized DFT calculation. In (b), the schematic diagram for the interaction of an adsorbed H atom with the rest atom is displayed. The filled (open) circles represent the full (partial) occupation of spin-up or -down electron, while the arrow indicates charge transfer. The top view of spin density is displayed in (c), together with the side view taken in the cross section along the dashed line. The majority (minority) spin density is displayed in bright (dark) color with an isosurface of 0.014 ($-$0.014) e/{\AA}$^3$. The numbers in (c) represent the spin moments (in ${\mu}_{\rm B}$) for the A$_1$, A$_2$, and A$_6$ atoms. The spin-polarized LDOS of A$_1$ and A$_2$ is given in (d). $E_{\rm ex}$ represents the exchange splitting of majority and minority bands. }
\end{figure}

\begin{figure}[htb]
\includegraphics[width=8.0cm]{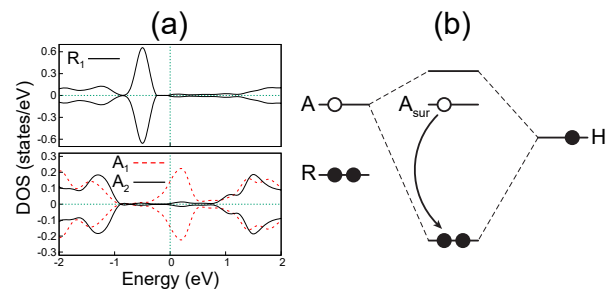}
\caption{ (Color online) (a) Calculated LDOS of the R$_1$, A$_1$, and A$_2$ atoms for the adsorption of a single H atom on the adatom A$_2$, obtained using the spin-unpolarized DFT calculation. In (b), the schematic diagram for the interaction of an adsorbed H atom with the adatom is displayed. The filled (open) circles represent the full (partial) occupation of spin-up or -down electron, while the arrow indicates charge transfer. A$_{\rm sur}$ represents the surrounding adatoms around A$_2$.}
\end{figure}

Due to the increased LDOS of A$_1$ and A$_2$ (A$_6$) at $E_{\rm F}$ [see Fig. 2(a)], the H adsorption on the R$_1$ atom may be expected to induce the local spin polarization around the adsorbed H atom. Indeed, our spin-polarized DFT calculation for the adsorption of a single H atom on the R$_1$ atom obtains the spin polarization of the surrounding A$_1$, A$_2$, and A$_6$ adatoms. Among several spin-polarized configurations (see Fig. S3 of the Supplemental Material), the most stable configuration is a ferrimagnetic spin ordering of A$_1$, A$_2$, and A$_6$ as shown in Fig. 2(c), where the spin directions of the two center adatoms A$_2$ and A$_6$ are the same with each other, but they are opposite to that of the corner adatom A$_1$. This ferrimangetic configuration is energetically favored over the spin-unpolarized one by 11 meV per 7$\times$7 unit cell. By integrating the spin density inside the PAW sphere with a radius of 1.574 {\AA} for Si, we obtain an identical spin moment of $m$ = 0.22 ${\mu}_{\rm B}$ for A$_2$ and A$_6$, while $m$ = $-$0.15 ${\mu}_{\rm B}$ for A$_1$. The relatively larger magnitude of spin moment for A$_2$ and A$_6$ may be associated with their higher LDOS at $E_{\rm F}$ [see Fig. 2(a)], which gives rise to a larger spin splitting. Interestingly, as shown in Fig. 2(c), the spin density is largely delocalized up to the fifth deeper atomic layers. Note that (i) the total spin moment $m_{\rm t}$ obtained including the interstitial region outside the PAW sphere is 0.66 ${\mu}_{\rm B}$ per 7$\times$7 unit cell and (ii) the sum of the spin moments of subsurface Si atoms is slightly larger than that of the A$_1$, A$_2$, and A$_6$ atoms. This large spin delocalization enables the spin coupling of the three neighboring adatoms (far-separated by ${\sim}$7.7 {\AA}) through the subsurface layers. In this sense, it is most likely that the H adsorption on R$_1$ induces the local spin polarizations at the nearest neighboring adatoms due to the intra-atomic exchange of localized DB electrons, and such spin moments are in turn coupled with each other via the interatomic exchange interaction. It is noteworthy that the local spin polarizations at A$_1$, A$_2$, and A$_6$ can be derived from Stoner's criterion $D(E_{\rm{F}})I > 1$~\cite{stoner}, where $D(E_{\rm{F}})$ is the LDOS at $E_{\rm F}$ and the Stoner parameter $I$ can be estimated dividing the exchange splitting $E_{\rm ex}$ by the corresponding magnetic moment [see Fig. 2(d)]~\cite{Burton11,Edwards06}. Here, the calculated values of $D(E_{\rm{F}})$ and $I$ are 0.203 and 6.2 (0.252 and 5.5) at A$_1$ (A$_2$), respectively, thereby satisfying $D(E_{\rm{F}})I > 1$. Interestingly, the present ferrimagnetic order of adatoms has two types of spin-spin interactions: i.e., the ferromagnetic (FM) coupling between A$_2$ and A$_6$, while the antiferromagnetic (AFM) coupling between A$_1$ and A$_2$ (or A$_6$). The FM coupling reduces the potential energy of repulsive electron-electron interactions due to the Pauli exclusion principle, whereas the AFM coupling lowers the kinetic energy of electrons. Therefore, the ferrimagnetic order of A$_1$, A$_2$, and A$_6$ is likely to involve the synergetic effects of the potential and kinetic energies in the three localized spins at the triangular sites.

Although the most stable H-adsorption site on Si(111)-(7$\times$7) is the rest-atom site, STM experiments identified the adatom site as a precursor or metastable intermediate state~\cite{prb2005,lo}. Using the spin-unpolarized DFT calculation, we optimize the structure of a single H atom adsorbed on the adatom site A$_2$, which is almost energetically degenerate (less than ${\sim}$3 meV per 7$\times$7 unit cell) with the adatom site A$_1$. We find that the H adsorption on the A$_2$ site is less stable than on the R$_1$ site by 0.187 eV per 7$\times$7 unit cell, in good agreement with the value (${\sim}$0.2 eV) of a previous DFT calculation~\cite{vittadini}. Further, based on the nudged elastic-band method~\cite{neb}, we calculate the energy profile along the H-diffusion pathway on going from the A$_1$ to the R$_1$ site, and obtain an energy barrier of 1.38 eV, consistent with that (1.3 eV) of a previous DFT calculation~\cite{vittadini}. As shown in Fig. 3(a), when H adsorbs on the A$_2$ site, the LDOS peak of A$_2$ disappears around $E_{\rm F}$. Meanwhile, the LDOS peak of A$_1$, located just above $E_{\rm F}$, slightly shifts toward lower binding energy, implying a charge transfer from A$_1$ to A$_2$. Note that the LDOS obtained at the rest atom R$_1$ remains intact compared to the case of a clean Si(111)-(7$\times$7) surface [see Figs. 3(a) and 1(c)]. Figure 3(b) shows the schematic diagram for the interaction of an adsorbed H atom with the A$_2$ atom. In order to make an Si$-$H bond formation, the A$_2$ atom needs 7/12 more electrons because it is partially occupied by 5/12 DB electrons before H adsorption. Therefore, unlike the above-discussed H adsorption on the R$_1$ site, H adsorption on the A$_2$ site needs to abstract electrons from surrounding adatoms. This charge abstraction decreases the LDOS of surrounding adatoms below $E_{\rm F}$ [see the LDOS of A$_1$ in Fig. 3(a)]. It is thus unlikely to show spin polarization according to Stoner's criterion. Indeed, our spin-polarized DFT calculation for the H adsorption on the adatom site A$_1$ or A$_2$ does not show any spin polarization.

It is interesting to note that, as the number of H atoms adsorbed on the adatom sites increases, the DB electrons abstracted from unreacted surrounding adatoms become insufficient to form the Si$-$H bonds. Hence, some neighboring rest atoms begin to participate in charge donation to the H-adsorbed adatoms, leading to the shift of the LDOS peaks of the rest atoms toward $E_{\rm F}$. This de-charge transfer may induce spin polarization at the rest atoms. Our spin-unpolarized DFT calculation shows that, when the number (denoted as $n_{\rm A}$) of adsorbed H atoms on the adatom sites increases to 5, the LDOS peak of R$_1$ is located near $E_{\rm F}$ [see Fig. 4(a)], thereby possibly inducing spin polarization due to the Stoner-type instability. Indeed, our spin-polarized DFT calculation demonstrates that the case of $n_{\rm A}$ = 5 exhibits a spin splitting in the LDOS of R$_1$ [see Fig. 4(b)], resulting in the spin polarization at R$_1$ [see Fig. 4(c)]. Figure 4(c) also displays the energetically most stable spin configurations for $n_{\rm A}$ = 6 (saturating the adatom sites in the F half cell) and 12 (saturating the adatoms sites in the F and UF half cells), respectively. Here, for $n_{\rm A}$ = 6 we obtain the AFM order of two rest atoms $R_1$ and $R_3$ with $m$ = ${\pm}$0.57 ${\mu}_{\rm B}$, while for $n_{\rm A}$ = 12 the ferrimagnetic order of six rest atoms in the F and UF unit cells [see each atomic spin moment in Fig. 4(c)]. It is thus likely that the number of adsorbed H atoms on the adatom sites can vary the pattern of spin polarization at the rest-atom sites. Similarly, we find that, when the numbers (denoted as $n_{\rm R}$) of adsorbed H atoms on the rest-atom sites are 2, 3, and 6, the ferrimagnetic orders of adatoms become most stabilized with different spin moments, as shown in Fig. 4(d).

\begin{figure}[htb]
\includegraphics[width=8.0cm]{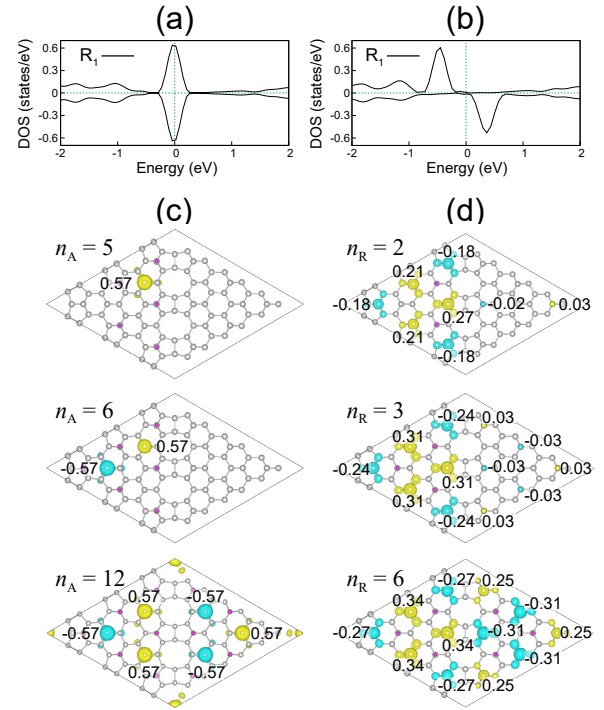}
\caption{ (Color online) (a) Spin-unpolarized and (b) spin-polarized LDOS of the R$_1$ atom for $n_{\rm A}$ = 5. Spin densities for $n_{\rm A}$ = 5, $n_{\rm A}$ = 6, and $n_{\rm A}$ = 12 are displayed in (c), while those for $n_{\rm R}$ = 2, $n_{\rm R}$ = 3, and $n_{\rm R}$ = 6 in (d). Here, the majority (minority) spin density is drawn with an isosurface of 0.014 ($-$0.014) e/{\AA}$^3$. In (c) and (d), the numbers represent the spin moments (in ${\mu}_{\rm B}$) for rest atoms and adatoms, respectively.
}
\end{figure}

\section{Summary}

Based on first-principles DFT calculations, we presented the spin orderings of Si DBs at the atomic scale, induced by H adsorption on the Si(111)-(7$\times$7) surface. We revealed that the adsorption of a single H atom on a rest-atom site causes a charge transfer back to the nearest neighboring adatoms, therefore giving rise to an increase in the LDOS of such three adatoms. The resulting high DOS at $E_{\rm F}$ is attributed to the Stoner-type instability to yield a ferrimagnetic order of adatom DBs around the adsorbed H atom. Meanwhile, when an H atom adsorbs on an adatom site, a charge abstraction from its surrounding adatoms takes place, forming an H-Si bond without spin polarization. However, as the number of adsorbed H atoms on the adatom sites within the 7$\times$7 unit cell increases, some neighboring rest atoms begin to participate in charge donation to the adatoms, therefore shifting the LDOS peak of such rest atoms toward $E_{\rm F}$ to induce a ferrimagnetic order. In the present study, we propose the underlying mechanism of the H-induced spin orderings of DBs on the Si(111)-(7$\times$7) surface in terms of the intra-atomic and interatomic exchanges of localized DB electrons through the subsurface layers. We anticipate that the present way to the design of nanoscale magnetism on the Si(111)-(7$\times$7) surface is rather generic and hence, it should be more broadly applicable to tailor the spin orderings of Si DBs by using other adsorbates.

\vspace{0.4cm}

\noindent {\bf Acknowledgement.}
We are grateful for discussions with Y. Jia. This work is supported by the NSFC of China (Grant Nos. 11804306, 11674289, and 11504332) and National Research Foundation of Korea (NRF) grant funded by the Korean Government (Grant Nos. 2015M3D1A1070639 and 2016K1A4A3914691). The calculations were performed at the High Performance Computational Center of Zhengzhou University. J.H.C acknowledges the strategic support program (KSC-2017-C3-0080) for supercomputing application research from the KISTI supercomputing center. \\

X.-Y. R. and C.-Y. N. contributed equally to this work.

\noindent $^{*}$ Corresponding author: chojh@hanyang.ac.kr

\end{document}